\documentclass[aps, prd, nofootinbib, preprintnumbers, showkeys, preprint]{revtex4}
\usepackage{amsmath}
\usepackage{amssymb}
\allowdisplaybreaks

\usepackage{makeidx}
\usepackage{amsfonts}
\usepackage[usenames,dvipsnames]{pstricks}
\usepackage{subfigure}
\usepackage{epsfig}
\usepackage{pst-grad} 
\usepackage{mathtools}
\usepackage[colorlinks,hyperindex]{hyperref}
\usepackage{array}
\usepackage{tikz}
\usetikzlibrary{decorations.markings,math,shapes.misc,patterns,arrows.meta,calc,patterns,angles,quotes}

\hypersetup
{	colorlinks,%
	citecolor=black,%
	linkcolor=black,%
	urlcolor=black,%
}

\usepackage{allpurpose}

\newcommand {\sign}{\mathrm{sign}}

\begin{document}

\title{Time delay in the strong field limit for null and timelike signals and its simple interpretation}

\author{Haotian Liu}
\affiliation{MOE Key Laboratory of Artificial Micro- and Nano-structures, School of Physics and Technology, Wuhan University, Wuhan, 430072, China}

\author{Junji Jia}
\email{junjijia@whu.edu.cn}
\affiliation{MOE Key Laboratory of Artificial Micro- and Nano-structures, School of Physics and Technology, Wuhan University, Wuhan, 430072, China}


\date{\today}

\begin{abstract}
Gravitational lensing can happen not only for null signal but also timelike signals such as neutrinos and massive gravitational waves in some theories beyond GR. In this work we study the time delay between different relativistic images formed by signals with arbitrary asymptotic velocity $v$  in general static and spherically symmetric spacetimes. A perturbative method is used to calculate the total travel time in the strong field limit, which is found to be in quasi-series of the small parameter $a=1-b_c/b$ where $b$ is the impact parameter and $b_c$ is its critical value. The coefficients of the series are completely fixed by the behaviour of the metric functions near the particle sphere $r_c$ and only the first term of the series contains a weak logarithmic divergence. The time delay $\Delta t_{n,m}$ to the leading non-trivial order was shown to equal the particle sphere circumference divided by the local signal velocity and multiplied by the winding number and the redshift factor.  By assuming the Sgr A* supermassive black hole is a Hayward one, we were able to validate the quasi-series form of the total time, and reveal the effects of the spacetime parameter $l$, the signal velocity $v$ and the source/detector coordinate difference $\Delta\phi_{sd}$ on the time delay. It is found that as $l$ increase from 0 to its critical value $l_c$, both $r_c$ and $\Delta t_{n,m}$ decrease. The variation of $\Delta t_{n+1,n}$ for $l$ from 0 to $l_c$ can be as large as $7.2\times 10^1$ [s], whose measurement then can be used to constrain the value of $l$. While for ultra-relativistic neutrino or gravitational wave, the variation of $\Delta t_{n,m}$ is too small to be resolved. The dependence of $\Delta t_{n,-n}$ on $\Delta \phi_{sd}$ shows that to temporally resolve the two sequences of images from opposite sides of the lens,  $|\Delta \phi_{sd}-\pi|$  has to be larger than certain value, or equivalently if $|\Delta \phi_{sd}-\pi|$ is small, the time resolution of the observatories has to be large. 
\end{abstract}

\keywords{Gravitational lensing; time delay; strong field limit; neutrino; timelike signal}

\maketitle

\section{Introduction}

Deflection of light signal near massive celestial bodies was the most convincing phenomenon that helped establishing General Relativity as the correct description of gravity \cite{Dyson:1920cwa}. Nowdays deflection of signals by gravity is almost exclusively used in gravitational lensing (GL), which has become a powerful tool in astrophysics and cosmology. GL has been used to constrain not only properties of the lens, such as the mass distribution of galaxy \cite{Bradac:2006er,Treu:2010uj,Suyu:2012aa}, the structure of dark matter halos \cite{Guo:2009fn,Zu:2015cpa,Wechsler:2018pic}, accretion of materials \cite{Morgan:2010xf,Diemer:2014xya}, but also those of the source, such as supernova explosion mechanism \cite{Kelly:2014mwa,Goobar:2016uuf}. Observables in GL can also correlate with properties of signal particles forming the GL images \cite{Fan:2016swi}. 

Traditionally, GL observations were always done using light signals of various wavelength. However, with the discovery of neutrinos form SN1987A \cite{Hirata:1987hu,Bionta:1987qt}  and more recently from blazer TXS
0506 + 056 \cite{IceCube:2018dnn,IceCube:2018cha}, and the discovery of gravitational waves (GW) due to binary black hole (BH)/neutron star mergers \cite{Abbott:2016blz,Abbott:2016nmj,Abbott:2017oio}, especially the GBR+GW dual observation \cite{TheLIGOScientific:2017qsa}, it is clear that neutrinos and GW can also act as messengers to have GL effects. 

In (strong) GL observations, the apparent angle of the images and the time delay between them are two most important observables that were widely used to reveal information about the lens, source and messenger. For both neutrino and GW observatories, currently their angular resolution are both too low to distinguish different images. However, the time measurement for these signals are usually very precise, reaching $\mathcal{O}(1)$ [ns] for neutrino events \cite{Eberhardt:2017vce,Djurcic:2015vqa}
and $\mathcal{O}(1)$ [ms] for GW observations \cite{TheLIGOScientific:2017qsa}. Therefore theoretical study and corresponding observations on time delay in GLs of these signals might bear fruit earlier than  resolving different images formed by them. 

In this work, we study time delay of timelike and null signals with general asymptotic velocity in the strong field limit (SFL) in static and spherically symmetric (SSS) spacetimes. In this limit, the signal's trajectory approaches the critical particle/photon sphere and the signal might loop around the central lens many time before reaching the detector, forming series of relativistic images from each side of the lens.  The time delay of light signal in SSS spacetimes in this limit has been studied by Bozza in Ref. \cite{Bozza:2003cp} as a distance estimator and then followed by many works in particular spacetimes or gravitational theories \cite{Eiroa:2004gh,Eiroa:2005ag,Eiroa:2013nra,Cavalcanti:2016mbe,Zhao:2017cwk,Wang:2019cuf, Sahu:2015dea,Man:2015iga, Majumdar:2005ba,Wang:2019cuf, Gyulchev:2008ff,Sahu:2013uya}. In this work, we not only extend it to arbitrary signal velocity, but develop a trackable way to calculate the total travel time and time delay to any desired order, which was never done before. Moreover, we also show that the time delay is given by a simple formula, Eq. \eqref{eq:tdtotapp}, allowing a very simple and intuitive understanding, i.e., Eq. \eqref{eq:tdsimpund}.

The work is organized as follows. In Sec. \ref{sec:perthe}, we develop the perturbative method used for the computation of the total travel time in the SFL. We will show that the total travel time takes a quasi-series form, which is then used in Sec. \ref{sec:tdinsfl} to find the time delay between different images.  It is show that the time delay equals to the circumference of the particle sphere divided by the local velocity of the signal and then multiplied by the redshift factor and the winding number. 
In Sec. \ref{sec:harcase}, we then apply these results to the Hayward BH spacetime and study the dependance of the time delay on the spacetime charge parameter $l$ and signal velocity $v$. The result reveals that the time delay can be used to constrain $l$ quite well but not $v$ since for both supernova neutrinos and GWs, their speeds have been well constrained to be extremely close to light speed. 

\section{The Perturbation method\label{sec:perthe}} 

The perturbative method we used here to calculate the total travel time is adopted from Ref. \cite{Jia:2020huang} which was to calculate the deflection angle in the SFL, and here this method is used to a different integral. Therefore in this section, we will first recap the essential steps to help to understand the method, and then apply the method to the suitable integral defining the total travel time of the signal. 

We start from the most general SSS metric described by
\be
\dd s^2=-A(r) \dd t^2 + B(r) \dd r^2 + C(r) (\dd \theta ^2 + \sin^2 \theta  \dd \varphi^2), \label{eq:sssmetric}
\ee
where $(t,~r,~\theta ,~\varphi)$ are coordinates and $A,~B,~C$ are metric functions depending on $r$ only. The corresponding geodesic equations can always be transformed to the equatorial plane $( \theta  = \pi/2)$, and then become
\begin{align}
  \dot{r}^2 & = \frac{\lb \frac{E^2}{A(r)}-\kappa \rb C(r) - L^2 }{B(r) C(r)}, \label{eq:drodk} \\
  \dot{t} & = \frac{E}{A(r)}, \label{eq:dtodk} \\
  \dot{\varphi} &  = \frac{L}{C(r)}, \label{eq:dphiodk}
\end{align}
where $\kappa=0,~1$ for massless and massive signals and $L,~E$ are the angular momentum and energy of the signal (per unit mass). $L,~E$ can relate to the impact parameter $b$ and the asymptotic velocity $v$ of the signal by
\be
L=\frac{bv}{\sqrt{1-v^2}},~E=\frac{1}{\sqrt{1-v^2}}. \label{eq:letob}
\ee

The travel time from a source at $r=r_s$ to a detector at $r=r_d$ (see Fig. \ref{fig:lensfig}), after using Eqs.  \eqref{eq:drodk} and \eqref{eq:dtodk}, is then 
\be
t=\sum_{i=s,d} \int_{r_0}^{r_i} \frac{\dot{t}}{\dot{r}}\dd r=\sum_{i=s,d} \int_{r_0}^{r_i} \frac{E \sqrt{B(r)C(r)}}{L A(r)}\cdot\frac{L}{\sqrt{\lb \frac{E^2}{A(r)}-\kappa \rb C(r) - L^2 } } \dd r, \label{eq:ttint}
\ee
where $r_0$ is the closest approach of the trajectory to the lens. According to Eq. \eqref{eq:drodk}, $r_0$ satisfies $\dot{r}|_{r=r_0}=0$, i.e.,
\be
\lb \frac{E^2}{A(r_0)}-\kappa \rb C(r_0) - L^2 = 0. \label{eq:rztol}
\ee
Together with Eq. \eqref{eq:letob}, this provides a relation connecting the closest approach $r_0$ and the impact parameter $b$
\be
\frac{1}{b}=\frac{\sqrt{E^2-\kappa}}{L}=\sqrt{\frac{E^2-\kappa}{\lb \frac{E^2}{A(r_0)} - \kappa \rb C(r_0) }}. \label{eq:defineb}
\ee

On the other hand, when $r_0$ or $b$ is small enough, the signal will spiral into a compact sphere, the particle (or photon) sphere, and not escape back to infinity. The radius $r_c$ of this sphere is defined as the critical point of the denominator of Eq. \eqref{eq:ttint}
\be
\dd \lsb \lb \frac{E^2}{A(r)}-\kappa \rb C(r) \rsb/ \dd r |_{r=r_c}=0. \label{eq:definerc}
\ee
If  $r_0$ approaches $r_c$ from above, we call the corresponding $b$ the critical impact parameter  and denote it by $b_c$. Using Eq. \eqref{eq:defineb}, it is related to $r_c$ by
\be
\frac{1}{b_c}=\sqrt{\frac{E^2-\kappa}{\lb \frac{E^2}{A(r_c)} - \kappa \rb C(r_c) }}. \label{eq:definebc}
\ee

The total travel time \eqref{eq:ttint} in the SFL is difficult to compute analytically even for the simpler SSS spacetimes and therefore approximation methods are desired. In Ref. \cite{Jia:2020huang} we have developed a perturbative way to expand a similar integrand for the computation of signal's deflection angle. Here we adopt that method to the computation of the total travel time in the SFL. First, we define a function $p(x)$ inspired by Eq. \eqref{eq:defineb} as
\be
p\lb x \rb = \frac{1}{b_c} - \sqrt{ \frac{E^2 - \kappa }{ \lb \frac{E^2}{A\lb 1/x \rb} - \kappa \rb C\lb 1/x\rb } }, \label{eq:definepx}
\ee
and define its inverse function as $q(x)$. From Eq. \eqref{eq:defineb}, it is clear that 
\be p\lb \frac{1}{r_0}\rb=\frac{1}{b_c}-\frac{1}{b}\equiv \frac{a}{b_c}, ~\text{where}~a\equiv 1-\frac{b_c}{b}\label{eq:adef}\ee
so that taking its inverse function we have 
\be \frac{1}{r_0} = q\lb \frac{1}{b_c} - \frac{1}{b} \rb = q\lb \frac{1-b_c/b}{b_c} \rb. \label{eq:qrel} \ee

We then can do the following change of variable in Eq. \eqref{eq:ttint} from $r$ to $\xi$ connected by function $q$ or $p$
\be
\frac{1}{r}=q\lb \frac{\xi}{b_c}\rb,~\text{i.e.,}~p\lb \frac{1}{r}\rb=\frac{\xi}{b_c}. \label{eq:pqtorxi}
\ee
Note that although the analytical form of $p(x)$ is clear once the metric function is given, the inversion process to find $q(x)$ is not always possible analytically. Fortunately, what is required in our later computation is the series expansion of $q(x)$ and it can always be worked out using the Lagrange inverse theorem from the series form of $p(x)$. 
Under this change of variable, and noticing Eqs. \eqref{eq:definepx}, \eqref{eq:adef}, the integration limits and various factors in the integrand of Eq. \eqref{eq:ttint} are changed according to 
\begin{subequations} \label{eq:varinttfrtoxi}
\begin{align}
 & r_0 \to 1-\frac{b_c}{b}= a,~r_i \to b_c p\lb \frac{1}{r_i}\rb\equiv \eta_i,~i=s,d, \label{subeq:relbtoa}  \\
 & \frac{L}{\sqrt{\lb \frac{E^2}{A(r)}-\kappa\rb C(r)-L^2}} \to \frac{1-\xi}{\sqrt{\lb \frac{b_c}{b}\rb^2 - \lb \xi-1\rb^2 } },  \\
 & \frac{E \sqrt{B(r)C(r)}}{L A(r)} \to \frac{\sqrt{B(1/q)C(1/q)}}{bvA(1/q)}, \\
 & \dd r \to -\frac{q'}{q^2} \frac{\dd \xi}{b_c},
\end{align}
\end{subequations}
where $q=q\lb \frac{\xi}{b_c}\rb$ and $q'$ is its derivative. Here the $\eta_{s,d}$ are nothing but the sine value of the apparent angles of the signal at the source and detector respectively [cite a few of our works]. In the SFL and large $r_{s,d}$ limits, we have $a\to 0^+$ and $\eta_{s,d}\to 1^-$ respectively. Grouping these terms together, we obtain the transformed total travel time as
\be
t=\sum_{i=s,d} \int_{a}^{\eta_i} \frac{\sqrt{B(1/q)C(1/q)}}{A(1/q)} \cdot \frac{\xi-1}{\sqrt{\lb 2-a-\xi\rb \lb \xi-a\rb } } \cdot \frac{(1-a)q'}{vb_c^2q^2} \dd \xi. \label{eq:ttintxi}
\ee
Note that this form of $t$ depends on the impact parameter $b$ through the parameter $a$, on the source/detector radius through $\eta_{s,d}$. Its dependence on all other parameters of the spacetime is through the metric functions and the critical impact parameter $b_c$, which also appears in $q=q(\xi/b_c)$.

\subsection{Perturbative expansion of total travel time} \label{subsec:tappone}

The beauty of the above change of variable \eqref{eq:pqtorxi} is that it transforms the infinite integration range to a finite range and allow the resultant integrand to be expanded in the small $\xi$ limit, so that an perturbative integration can be carried out.

To see how the expansion is carried out, we first split the integrand into two factors, $\frac{1}{\sqrt{\xi-a}}$ and $y(\xi)$ with
\be
y(\xi)=\frac{\sqrt{B(1/q)C(1/q)}}{A(1/q)} \cdot \frac{\xi-1}{\sqrt{2-a-\xi} } \cdot \frac{(1-a)q'}{vb_c^2q^2}. \label{eq:yformori}
\ee
The factor $\frac{1}{\sqrt{\xi-a}}$ will be directly integrated later while the factor $y(\xi)$ should be further treated. Since in the SFL, the main part of the total time is contributed from the integration near small $\xi$, we can further split $y(\xi)$ into two factors and expand them in the small $\xi$ limit, i.e., 
\begin{subequations} \label{eq:ttyexptwop}
\begin{align}
\frac{1}{\sqrt{2-a-\xi} }=&\sum_{n=0}^{\infty} \frac{(2n-1)!!}{(2n)!!}\frac{\xi^n}{(2-a)^{n+\frac{1}{2}}}, \label{eqs:ttyexpp1} \\
\frac{\sqrt{B(1/q)C(1/q)}}{A(1/q)} \frac{ \lb \xi-1 \rb (1-a)q'}{vb_c^2q^2} =&(1-a) \sum_{n=-1}^{\infty} f_n \xi^{\frac{n}{2}}, \label{eqs:ttyexpp2}
\end{align}
\end{subequations}
where in Eq. \eqref{eqs:ttyexpp2} the index starting from $-1$ because of the $q'$ term, and the $f_n$ are the expansion coefficients that can be worked out once the metric functions are known. We point out that it is in expansion \eqref{eqs:ttyexpp2} that the series form of $q(\xi/b_c)$ is needed and can be obtained using the Lagrange inverse theorem from the series expansion of $p(x)$. We also emphasis that the coefficients $f_n$ will not depend on the initial/boundary conditions of the trajectory, such as the impact parameter $b$ and $r_{s,d}$, but only on the metric functions and spacetime parameters therein. 
 
Collecting these expansions according to the powers of $\xi$, we see that $y(\xi)$ becomes 
\be
y(\xi)=\sum_{n=-1}^{\infty} \frac{1-a}{\lb 2-a\rb^{[\frac{n+1}{2}]+\frac{1}{2}}} \sum_{m=0}^{[\frac{n+1}{2}]} y_{n,m} a^m \xi^{n/2}, \label{eq:yexpform}
\ee
where a sum over a finite terms of $a^m$ appears because different powers of the denominator factor $(2-a)$ in Eq. \eqref{eqs:ttyexpp1} mix into the coefficient of the same $\xi^{n/2}$ power. The coefficients $y_{n,m}$ can be obtained from the coefficients $f_n$ and other factors in Eqs. \eqref{eq:ttyexptwop} but their exact forms are too tedious to show here. 
Further substituting Eq. \eqref{eq:yexpform} into the total time, we finally get
\be
t=\sum_{i=s,d} \sum_{n=-1}^{\infty} \sum_{m=0}^{[\frac{n+1}{2}]} \frac{1-a}{\lb 2-a\rb^{[\frac{n+1}{2}]+\frac{1}{2}}} y_{n,m} a^m \int_{a}^{\eta_i} \frac{\xi^{n/2}}{\sqrt{\xi-a}}\dd \xi. \label{eq:ttgenexpform}
\ee
The integrability of this formula relies on the integration of the form $\displaystyle \int_{a}^{\eta_i} \frac{\xi^{n/2}}{\sqrt{\xi-a}}\dd \xi$. Fortunately, this kind of integrals can always be worked out for integers $n$, and the results are given in Eq. \eqref{eq:ttgenintm} in Appendix \ref{app:hoyz}. Using these results, the total travel time $t$ becomes
\begin{align}
t=\sum_{i=s,d} \sum_{k=0}^{\infty} \sum_{m=0}^{k} &\frac{\lb 1-a\rb a^m}{\lb 2-a\rb^{k+\frac{1}{2}}} \lcb y_{2k-1,m} \cdot \frac{a^kC_{2k}^k}{4^k} \Bigg[  -\ln a +2 \ln\lb \sqrt{\eta_i}+\sqrt{\eta_i-a}\rb  \right.  \nonumber \\
 & \left. \left. + \sum_{j=1}^{k}\frac{ 4^j \eta_i^j }{j a^j C_{2j}^{j}} \sqrt{1-\frac{a}{\eta_i}}\rsb + y_{2k,m} \cdot \sum_{j=0}^{k} \frac{2 C_{k}^{j}a^{k-j}\lb \eta_i-a\rb^{j+1/2}}{2j+1} \rcb. \label{eq:ttgentotform}
\end{align}
Here the first and second terms in the curl bracket are due to the integration of odd and even powers of $\xi^{n/2}$ respectively. It is also important to notice that all the dependence of $t$ on $a$ and $\eta_{s,d}$ in this formula has been shown explicitly, and the coefficients $y_{n,m}$ only contain the signal kinetic parameter $v$ and spacetime parameters through $b_c$.

It is not difficult to notice that all the functions involved in $t$ are quite elementary. Because of this, in the SFL (i.e, $b\to b_c^+,~a\to 0^+$), it can be further expanded into a quasi-series of small $a$, which yield a  from 
\be 
t=\sum_{k=0}^\infty \lsb C_k \ln a+D_k(\eta_s,\eta_d)\rsb a^k. 
\label{eq:totaltaf}\ee
where in the coefficient of each order of $a^k$, there is  only one term that contains $\ln a$. The coefficients $C_k$ and $D_k$ can be worked out from Eq. \eqref{eq:ttgentotform} and it is seen that only the $D_k$'s (but not the $C_k$'s) depend on $\eta_{s,d}$. This total time \eqref{eq:totaltaf} resembles  the same form as the deflection angle in the SFL in Ref. \cite{Jia:2020huang}, although their coefficients $C_k$ and $D_k$ will be different. When the source and detector radius $r_{s,d}$ are not infinite, then in the $a\to 0$ limit, there is only one divergent term in $t$ contributed by the $\mathcal{O}(a)^0$ order. 
To the $\mathcal{O} \lb a \rb^0$ order, one can also directly recognize from Eq. \eqref{eq:ttgentotform} that only the $m=k=0$ terms contribute. The result to this order is
\begin{align}
 t (r_s,r_d,b) =&  \sum_{i=s,d}\lcb - \frac{\sqrt{2}}{2} y_{-1,0} \ln a + \frac{\sqrt{2}}{2} y_{-1,0} \ln \lb 4 \eta_i \rb + \sum_{n=0}^{\infty} \frac{y_{n,0}}{2^{\lsb \frac{n+1}{2}\rsb +\frac{1}{2}}} \cdot \frac{2 \eta_i^{\frac{n+1}{2}}}{n+1}  \rcb + \mathcal{O}\lb a\rb^1.  \label{eq:ttgentotformexpa}\\
= &C_0\ln a +D_0(\eta_s,\eta_d)+\mathcal{O}\lb a\rb^1 \label{eq:ttgentotformexpasim}
\end{align}
from which we can read off the coefficients $C_0$ and $D_0$ in Eq. \eqref{eq:totaltaf} as
\begin{align}
C_0 =&  - \sqrt{2} y_{-1,0} , \label{eq:c0def}\\
D_0(\eta_s,\eta_d)=&\sum_{i=s,d}\lsb  \frac{\sqrt{2}}{2} y_{-1,0} \ln \lb 4 \eta_i \rb + \sum_{n=0}^{\infty} \frac{y_{n,0}}{2^{\lsb \frac{n+1}{2}\rsb +\frac{1}{2}}} \cdot \frac{2 \eta_i^{\frac{n+1}{2}}}{n+1}  \rsb. 
\end{align}

Later on, we will use the total time \eqref{eq:ttgentotformexpa} to calculate the time delay between different relativistic images in the SFL. Different images correspond to different impact parameters and consequently, different $a$, but their $r_{s,d}$ and other spacetime parameters are exactly the same. This also implies that when subtracting two total times along the trajectories with slightly different $b$, the $D_0$ term will not contribute to the time delay at this order. 

\subsection{Computing coefficients $y_{n,m}$}

From the relation \eqref{eq:qrel} and the change of variable \eqref{eq:pqtorxi} we knew that the $\xi\to 0^+$ limit is also the $r_0\to r_c^+,~b\to b_c^+$ limit. Therefore the expansions \eqref{eq:ttyexptwop} at small $\xi$ or equivalently the coefficients $y_{n,m}$  should also be determined from the series expansion of the metric functions at $r=r_c$. Assuming these expansions are
\begin{subequations} \label{eqs:metfuncabc}
\begin{align}
A(r\to r_c)=& \sum_{n=0}^{\infty} a_n (r-r_c)^n, \\
B(r\to r_c)=& \sum_{n=0}^{\infty} b_n (r-r_c)^n, \\
C(r\to r_c)=& \sum_{n=0}^{\infty} c_n (r-r_c)^n, 
\end{align}
\end{subequations}
where $a_i,~b_i$ and $c_i$ are the coefficients, 
then the very definition of $r_c$ in Eq. \eqref{eq:definerc} becomes a constraint between the first few coefficients
\be
c_1\lb \frac{E^2}{a_0} - \kappa \rb = \frac{a_1 c_0 E^2}{a_0^2}. \label{eq:genconrc}
\ee
The impact parameter in Eq. \refer{eq:definebc} can also be expressed using $a_0$ and $c_0$ as
\be
b_c=\frac{1}{v} \sqrt{\frac{c_0}{a_0} \lb 1 - a_0 + v^2 a_0 \rb }. \label{eq:bcinff}
\ee

Using these metric expansions and going through the process from Eqs. \eqref{eq:yformori} to \eqref{eq:yexpform}, the $y_{n,m}$'s can be computed. In particular, the first two $y_{n,0}$ are found to be
\begin{subequations} \label{eqs:gencoeyaz}
\begin{align}
y_{-1,0}=&\frac{1}{\sqrt{2}v a_0} \sqrt{\frac{b_0 c_0}{T_2} }, \\
y_{0,0} =& -\frac{b_c \lsb \lb 2 a_1 b_0 c_0 - a_0 b_1 c_0 -a_0 b_0 c_1\rb T_2 + 2 a_0 b_0 c_0 T_3 \rsb }{2 v a_0^2 \sqrt{b_0 c_0} T_2^2 }, 
\end{align}
\end{subequations}
where
\begin{subequations} \label{eqs:gencoeycoet}
\begin{align}
T_2=& \frac{1}{v^2 a_0} \lsb c_2 \lb 1 - a_0 + a_0 v^2 \rb - \frac{ a_1 c_1 + a_2 c_0 }{a_0} + \frac{a_1^2 c_0}{a_0^2} \rsb , \\
T_3=& \frac{1}{v^2 a_0} \lsb c_3 \lb 1 - a_0 + a_0 v^2 \rb - \frac{ a_1 c_2 + a_2 c_1 + a_3 c_0 }{a_0} + \frac{a_1^2 c_1 + 2 a_1 a_2 c_0 }{a_0^2} - \frac{a_1^3 c_0 }{a_0^3} \rsb.
\end{align}
\end{subequations}
Higher order $y_{n,0}~(n>0)$ and $y_{n,m}~(m>0)$ can also be readily computed but are too long to present here. However from these higher orders, we are able to assert that for general $n$, $y_{n,0}$ is determined by the metric expansion coefficients up to $a_{n+3}$, $b_{n+1}$ and $c_{n+3}$. 

\section{Time delay in the SFL} \label{sec:tdinsfl}

\begin{figure}[htp]
\begin{center}
\includegraphics[width=0.48\textwidth]{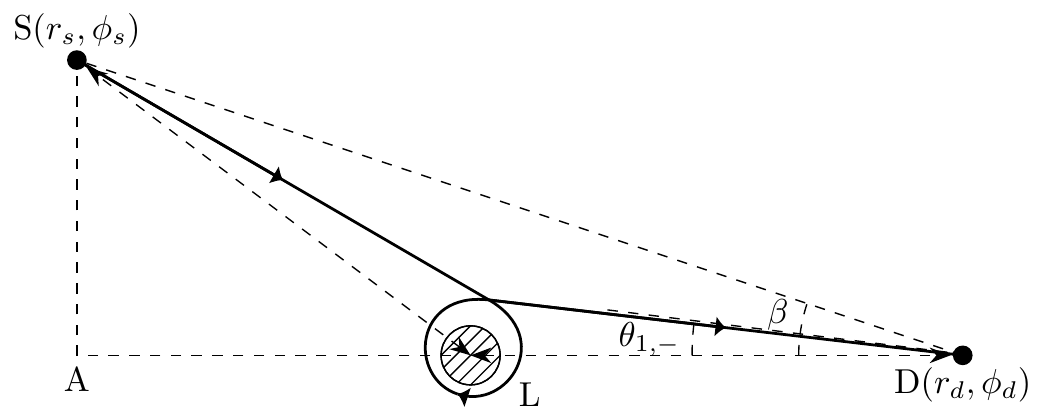}
\end{center}
\caption{The GL in the SFL. $(r_d,\phi_d)$ and $(r_s,\phi_s)$ are respectively the radial and angular coordinates of the detector D and source S. The signal winds clockwisely one loop around the lens L. $b$ is the impact parameter and the shadow stands for the BH.   } \label{fig:lensfig}
\end{figure}

In time measurement in GL, what is measured is not the total travel time but the time delay between different images of the source. For GL in the SFL, a typical path is illustrated in Fig. \ref{fig:lensfig}. There will exist two basic kinds of time delay: the time delay between signals from the same side of the lens but with different winding numbers around the center, and the time delay between signals from different sides of the lens but with the same winding number around the center. A general time delay, which we denote as $\Delta t_{n,m}~(n,m\in\mathbb{Z})$ to represent the time delay between images winding around the center counter-clockwisely $n$ and $m$ times, should be a combination of the former two basic time delays. Using Eq. \eqref{eq:ttgentotformexpasim}, this time difference can be written as 
\be
\Delta t_{n,m}=t(r_s,r_d,b_n)-t(r_s,r_d,b_m)=C_0\ln\lsb  \frac{1-b_c/b_n}{1-b_c/b_m}\rsb +\mathcal{O}(a)^1, \label{eq:tdingenforma}
\ee
where $b_n$ and $b_m$ are the impact parameters of the two images. Note the $D_0a^0$ term in Eq. \eqref{eq:ttgentotformexpasim} dose not affect the time delays since it is independent of $b$ and therefore the same for all trajectories. Furthermore, here $n,~m$ could be negative integers if the winding is indeed clockwise (although $b_n$ and $b_m$ are always positive). Note that the case that winding dose not actually happen corresponds to the weak field limit of the time delay and was considered in our previous work \cite{Liu:2020wcu}. 

Clearly, in order to compute the time delay, we shall find the corresponding impact parameters for the images in the SFL first. 
For this purpose, we will directly use the result from Eq. (38) of Ref. \cite{Jia:2020huang},
\be \displaystyle
b_{n} = \frac{b_c}{ 1-\mathrm{Exp}\lsb \displaystyle \frac{-\lb 2|n|+1+\sign(n) \rb \pi+D_{a0}+\sign(n) \Delta \phi_{sd} }{-C_{a0}} \rsb }, \label{eq:impbnpmform}
\ee
where $\Delta \phi_{sd}=\phi_s-\phi_d$ is the difference between the angular coordinates  $\phi_s$ of the source and $\phi_d$ of the detector (see Fig. \ref{fig:lensfig}). The coefficient \be 
C_{a0}=-b_c\sqrt{b_0/(c_0T_2)},\label{eq:ca0def}\ee
while the coefficient $D_{a0}$ is given in Eq. (28) of Ref.  \cite{Jia:2020huang} and dose not contribute to the time delay, as we will show in Eq. \eqref{eq:tdtotappori}. 

Now substituting Eq. \eqref{eq:impbnpmform} into   \eqref{eq:tdingenforma} and  after a small simplification we have
\bea
\Delta t_{n,m}&=&\frac{2\pi C_0}{C_{a0}}\lcb  \lb|n|- |m|\rb +\frac{ \lsb\sign(n)-\sign(m)\rsb (\pi- \Delta \phi_{sd})}{2\pi} \rcb \label{eq:tdtotappori}\\
&=&\frac{2\pi\sqrt{c_0}}{\sqrt{1-a_0(1-v^2)}}\frac{1}{\sqrt{a_0}}  \lcb  \lb|n|- |m|\rb +\frac{ \lsb\sign(n)-\sign(m)\rsb (\pi- \Delta \phi_{sd})}{2\pi} \rcb
\label{eq:tdtotapp} \eea
where $C_0$ in Eq. \eqref{eq:c0def}, $C_{a0}$ in Eq. \eqref{eq:ca0def} and $b_c$ in Eq. \eqref{eq:bcinff} are substituted and simplified. 
It is seen that the time delay depends only on the following parameters: the metric expansion coefficients $a_0,~c_0$,  the asymptotic signal velocity $v$ and a angular factor determined by the number of loops $n,~m$ and angular coordinate difference $\Delta\phi_{sd}$, but not on the finite distance $r_i$ and $r_f$ of the source and detector. 

We emphasis that this time delay is a very general result: it applies to GL with general asymptotic velocity $v$, general source/detector angular coordinate difference $\Delta \phi_{sd}$, general SSS spacetime with a particle sphere and arbitrary $n$ and $m$. Setting $v=1$ in Eq. \eqref{eq:tdtotapp} reduces it to previously known result in Ref. \cite{Bozza:2003cp} (Eq. (40) and (41)), which concentrated on null signals. 

Although the source and detector in GL in the SFL are usually far from the BH center, one would expect however when the winding number of two signals $n$ and $m$ are both large (in this case, $n,~m\geq 1$ are enough), the time delay between them, observed by a far away observer, should be equivalent to the circumference $2\pi r_c$ of the particle sphere divided by local signal velocity $v_l$ and then multiplied by the difference of loops $\Delta l$ (not necessarily an integer) between the two paths, and finally the gravitational redshift factor $\gamma$ from the particle sphere to the detector. That is, it is natural to expect that 
\be
\Delta t_{n,m}=\frac{2\pi r_c}{v_l} \cdot \Delta l\cdot \gamma. \label{eq:tdsimpund} 
\ee
Here we show that indeed, the above is exactly the time delay result \eqref{eq:tdtotapp}, which is found from more rigorous and lengthy calculations. 

For arbitrary SSS spacetime, it is always possible to choose the metric function $C(r)=r^2$ and therefore the its expansion coefficient at $r_c$ is $c_0=r_c^2$. That is, the particle sphere circumference $2\pi r_c=2\pi\sqrt{c_0}$, i.e., the numerator of the first factor in Eq.  \eqref{eq:tdtotapp}. For the local velocity $v_l$, in the SFL the signal circulates around the particle sphere, and therefore we only needs to consider the velocity due to angular motion. In an SSS spacetime,  this is given by
\be
v_l=\frac{r_c\dot{ \varphi}}{\gamma_{sr}}
\ee 
where $\gamma_{sr}=1/\sqrt{1-v_l^2}$ is the special relativity gamma factor. Then using Eqs. \eqref{eq:dphiodk}, \eqref{eq:letob} (setting $b$ to $b_c$), \eqref{eq:bcinff}, and $c_0=r_c^2$ sequencically, it is just an elementary algebra to solve the local velocity as
\be 
v_l=\sqrt{1-a_0(1-v^2)},
\ee
which is exactly the denominator of the first factor in Eq.  \eqref{eq:tdtotapp}. Thirdly, the difference in the number of loops $\Delta l$, after properly taking into account the opposite direction case, is simply the last factor of Eq. \eqref{eq:tdtotapp}. Lastly, in an asymptotically flat SSS spacetime described by metric \eqref{eq:sssmetric}, the gravitational redshift factor from $r_c$ to the detector which is located far away is simply $1/\sqrt{A(r_c)}=1/\sqrt{a_0}$, the second factor of Eq. \eqref{eq:tdtotapp}.
Grouping these factors together, therefore it is verified that for general SSS spacetime and timelike or null signal, the time delay in the SFL, Eq. \eqref{eq:tdtotapp}, has a very simple and intuitive understanding, Eq. \eqref{eq:tdsimpund}. 

\section{The Hayward BH spacetime case} \label{sec:harcase}

In this section, we apply our result to some particular spacetimes to check its validity, and examine the effect of $v$, $\Delta\phi_{sd}$, number of loops $n$ and $m$, and more importantly the spacetime parameters. The spacetime we study is the Hayward BH spacetime whose metric functions are \cite{Hayward:2005gi}
\be
A(r)=1-\frac{2Mr^2}{r^3+2l^2M},~B(r)=\frac{1}{A(r)},~C(r)=r^2, \label{eq:haymetform}
\ee
where $M$ is the spacetime mass and $l$ is the charge parameter. $|l|<4M/(3\sqrt{3}) \equiv l_c$ in order for the spacetime to be a BH one. Using Eqs. \eqref{eq:letob} and \eqref{eq:definerc}, the equation determining the particle sphere radius $r_c$ of this spacetime becomes
\begin{align}
4 l^4 M^2 v^2-8 l^2 M^2 \left(v^2-1\right)r_c^2 + 4 l^2 M v^2 r_c^3 + 4 M^2 \left(v^2-1\right)r_c^4+M \left(1-4 v^2\right)r_c^5+ v^2r_c^6 =0.
\end{align}
This is an six order polynomial of $r_c$ whose solution dose not have a closed algebraic form. However, after formally or numerically solving it, then substituting $r_c$ into the metrics \refer{eq:haymetform} and further into Eq. \refer{eq:definebc}, the critical impact parameter $b_c$ is found as
\be
b_c=\sqrt{\frac{2 l^2 M v^2r_c^2 -2 M \left(v^2-1\right)r_c^4 + v^2r_c^5}{v^2 \lb 2 l^2 M-2 Mr_c^2+r_c^3 \rb } }. \label{eq:haybcform}
\ee

To solve the $y_{n,m}$ that are needed in the total travel time \eqref{eq:ttgentotformexpa} and the time delay  \eqref{eq:tdtotapp}, then we should expand the metric functions at $r=r_c$ according to Eq. \eqref{eqs:metfuncabc}. The first few of these expansion coefficients are
\begin{subequations}\label{eqs:haymetabciform}
\begin{align}  
a_0=& \lb 1-\frac{2 M r_c^2}{2 l^2 M+r_c^3} \rb,~a_1 = -\frac{2 M \left(4 l^2 M r_c-r_c^4\right)}{\left(2 l^2 M+r_c^3\right)^2}, ~a_2=-\frac{2 M \left(4 l^4 M^2-14 l^2 M r_c^3+r_c^6\right)}{\left(2 l^2 M+r_c^3\right)^3}, \\
b_0=&\frac{2 l^2 M+r_c^3}{2 l^2 M-2 M r_c^2+r_c^3}, \\
c_0=&r_c^2,~c_1=2 r_c,~c_2=1.
\end{align}
\end{subequations}
Substituting them into Eqs. \eqref{eqs:gencoeyaz} and \eqref{eqs:gencoeycoet}, we can obtain the coefficients $y_{n,0}$ for the Hayward spacetime. The first two of them, denoted as $y_{-1,0,\mathrm{H}}$ and $y_{0,0,\mathrm{H}}$, are
\begin{subequations} \label{eqs:haymetyznform}
\begin{align}
y_{-1,0,\mathrm{H}}=&\frac{r_c}{v \sqrt{2 T_2} } \lsb \frac{ 2 l^2 M+r_c^3}{2 l^2 M+r_c^2 (r_c-2 M)} \rsb^{3/2}, \\
y_{0,0,\mathrm{H}}=&\frac{b_c\sqrt{2 l^2 M+r_c^3} }{v T_2^2 \lsb 2 l^2 M+r_c^2 (r_c-2 M)\rsb^{5/2} } \lcb 4 l^4 M^2 (T_2-r_c T_3)+4 l^2 M r_c^2 \lsb M (r_c T_3+2 T_2) \right. \right. \nonumber \\
  & \left. \left. +r_c (T_2-r_c T_3)\rsb +r_c^5 \lsb M (2 r_c T_3-5 T_2)+r_c (T_2-r_c T_3)\rsb \rcb ,
\end{align}
\end{subequations}
where
\begin{subequations} 
\begin{align}
T_2=&r_c^2 + \frac{2 M r_c^4}{v^2 \lsb 2 l^2 M-2 Mr_c^2+r_c^3\rsb}, \\
T_3=&\frac{1}{v^2 \lb 2 l^2 M-2 Mr_c^2+r_c^3 \rb^2} \lsb 8 v^2 l^4 M^2 r_c + 16 \left(1-v^2\right) l^2 M^2 r_c^3 + 8 v^2 l^2 M r_c^4 \right. \nonumber \\
  & \left. - 8 \left(1 - v^2\right) M^2 r_c^5 + 2 M \left(1-4 v^2\right) r_c^6 +2 v^2 r_c^7 \rsb.
\end{align}
\end{subequations}
High order $y_{n,0,\mathrm{H}} ~(n>0)$ can also be obtained by similar calculation but are too long to show here. 
The total travel time in the SFL in the Hayward spacetime can then be obtained from Eq. \eqref{eq:ttgentotformexpa} 
\be
t(r_s,r_d,b)= \sum_{i=s,d} \lcb- \frac{\sqrt{2}}{2} y_{-1,0,\mathrm{H}} \ln a + \frac{\sqrt{2}}{2} y_{-1,0,\mathrm{H}} \ln \lb 4 \eta_{i} \rb + \sum_{n=0}^{\infty} \frac{y_{n,0,\mathrm{H}}}{2^{\lsb \frac{n+1}{2}\rsb +\frac{1}{2}}} \cdot \frac{2 \eta_{i}^{\frac{n+1}{2}}}{n+1}\rcb. \label{eq:haytt}
\ee
Therefore, substituting metric expansion coefficients \eqref{eqs:haymetabciform} into Eq. \eqref{eq:tdtotapp}, the time delay in Hayward spacetime is simplified to
\begin{align}
\Delta t_{n,m,\mathrm{H}}=& \frac{2\pi r_c \left(2 l^2 M+r_c^3\right)}{2 l^2 M+r_c^2 (r_c-2 M)} \lsb \frac{2 M r_c^2}{2 l^2 M+r_c^2 (r_c-2 M)}+v^2 \rsb^{-\frac{1}{2}}  \nonumber \\
& \times \lcb  \lb|n|- |m|\rb +\frac{ \lsb\sign(n)-\sign(m)\rsb (\pi- \Delta \phi_{sd})}{2\pi} \rcb. \label{eq:dthayward}
\end{align}

This time delay depends on a few kinds of parameters: the spacetime parameters including its mass $M$ and charge $l$, the signal property --- its speed $v$, the winding numbers $n$ and $m$, and the source/detector angular coordinate difference $\Delta \phi_{sd}$ when $n$ and $m$ are not the same direction. Among these, the mass $M$ provides an overall scale for the time delay. As pointed out in Sec. \ref{sec:tdinsfl}, the first line of Eq. \eqref{eq:dthayward} actually is the time interval cost for the signal to loop one cycle around the particle/photon sphere.  In Schwarzschild spacetime, this time interval for light would be $2\pi r_{c,\mathrm{S}}/ \sqrt{1-2M/r_{c,\mathrm{S}}}=6\sqrt{3}M$ where $r_{c,\mathrm{S}}=3M$ for photon sphere in this spacetime. The charge $l$ is the main parameter characterizing this spacetime and $v$ is the parameter different from usual GL by light signal.  As $l$ deviates from zero or $v$ from 1, then this time interval also changes from the above value in a way dictated by the first line of Eq. \eqref{eq:dthayward}. 

\begin{center}
\begin{figure}[htp!]
\includegraphics[width=0.48\textwidth]{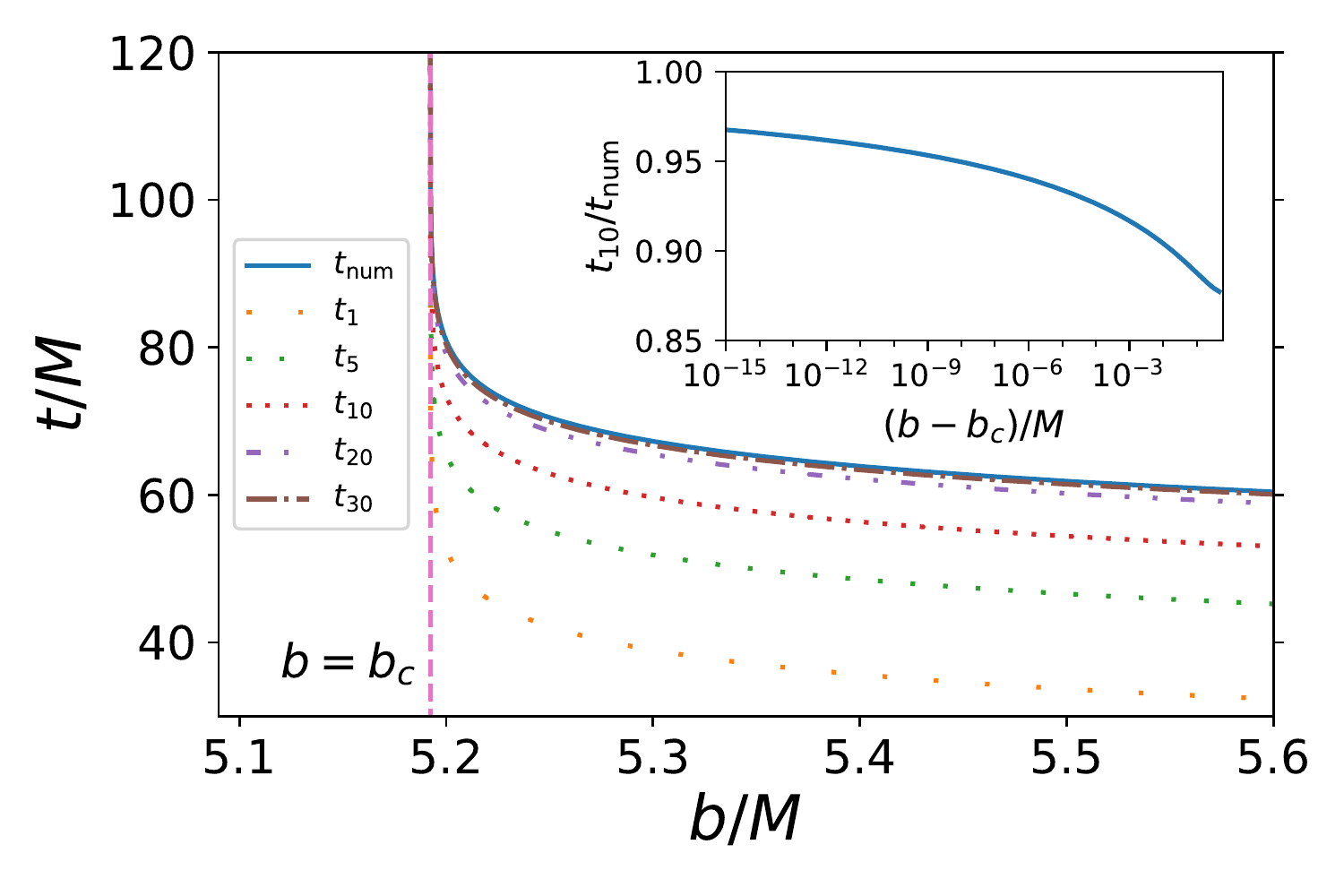}
\caption{Total travel time in Hayward spacetime as a function of $b$ using truncated series \eqref{eq:hayttm} (dashed or dotted lines) and the exact result using numerical integration of Eq. \eqref{eq:ttint} (the solid line). Inset: the ratio between the result truncated at order 10 and the numerical integral. } \label{fig:ttobrsd}
\end{figure}
\end{center}

To verify whether the time delay Eq. \eqref{eq:haytt} is accurate, we can first truncate the sum in its second term to order $m$ and define 
\be
t_m(r_s,r_d,b)= \sum_{i=s,d} \lcb- \frac{\sqrt{2}}{2} y_{-1,0,\mathrm{H}} \ln a + \frac{\sqrt{2}}{2} y_{-1,0,\mathrm{H}} \ln \lb 4 \eta_{i} \rb + \sum_{n=0}^{m} \frac{y_{n,0,\mathrm{H}}}{2^{\lsb \frac{n+1}{2}\rsb +\frac{1}{2}}} \cdot \frac{2 \eta_{i}^{\frac{n+1}{2}}}{n+1}\rcb, \label{eq:hayttm}
\ee
and then compare $t_m(r_s,r_d,b)$ with the total time $t_{\mathrm{num}}(r_s,r_d,b)$ obtained directly from numerical integration of Eq. \eqref{eq:ttint}. As long as the numerical integration is done to high enough precision,  $t_{\mathrm{num}}$ can be thought as the true value of the total travel time. 
In Fig. \ref{fig:ttobrsd}, we plot $t_1,~t_5,~t_{10},~t_{20},~t_{30}$ as well as $t_{\mathrm{num}}$ for $b$ around $b_c$, which is about $5.19M$ when we set $l=0.1M$ and $r_s=r_d=20M,~v=1$. It is seen that as the truncation order increases, the total time approaches its true value for all $b$ considered, and $t_{30}$ is non-distinguishable from $t_{\mathrm{num}}$ in the plot. Moreover, in the inset we see that for any fixed truncation order (we took $t_{10}$ as an example), the smaller the $b-b_c$, the better it approximates the true value of the travel time. 

\begin{center}
\begin{figure}[htp]
\subfigure[]{\label{fig:rcolv} \includegraphics[width=0.48\textwidth]{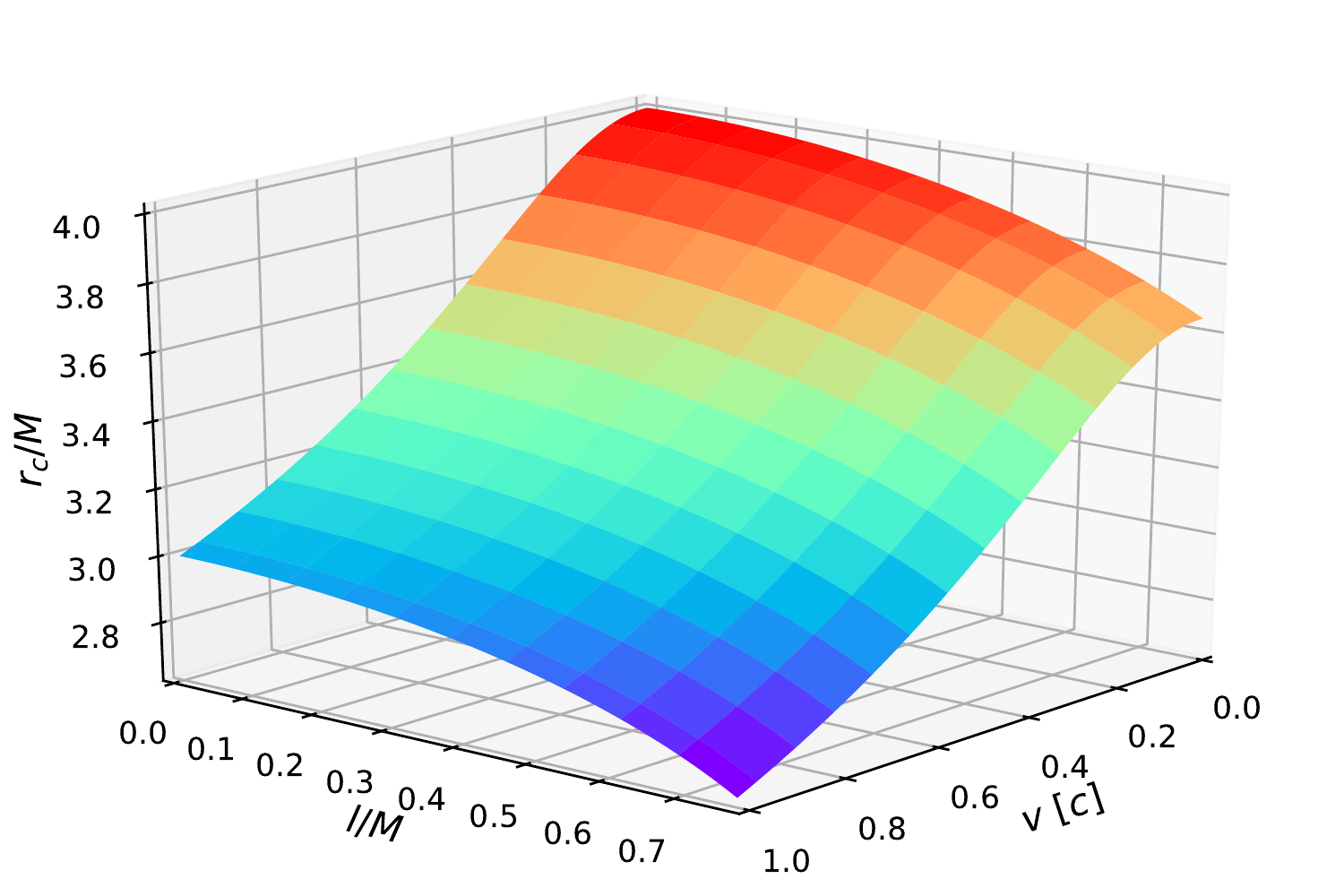}}\\
\subfigure[]{\label{fig:tdol} \includegraphics[width=0.48\textwidth]{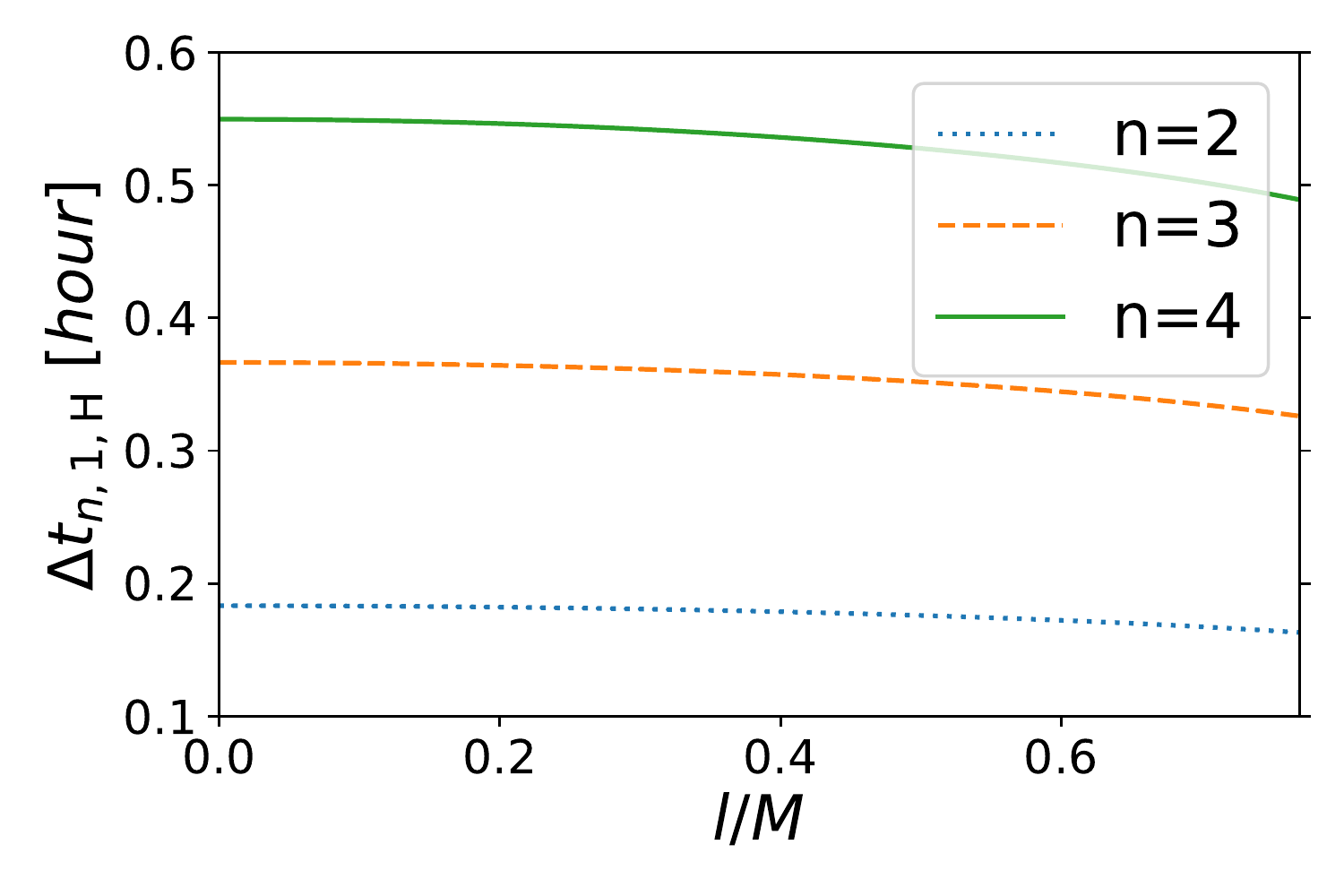}}
\subfigure[]{\label{fig:tdov} \includegraphics[width=0.48\textwidth]{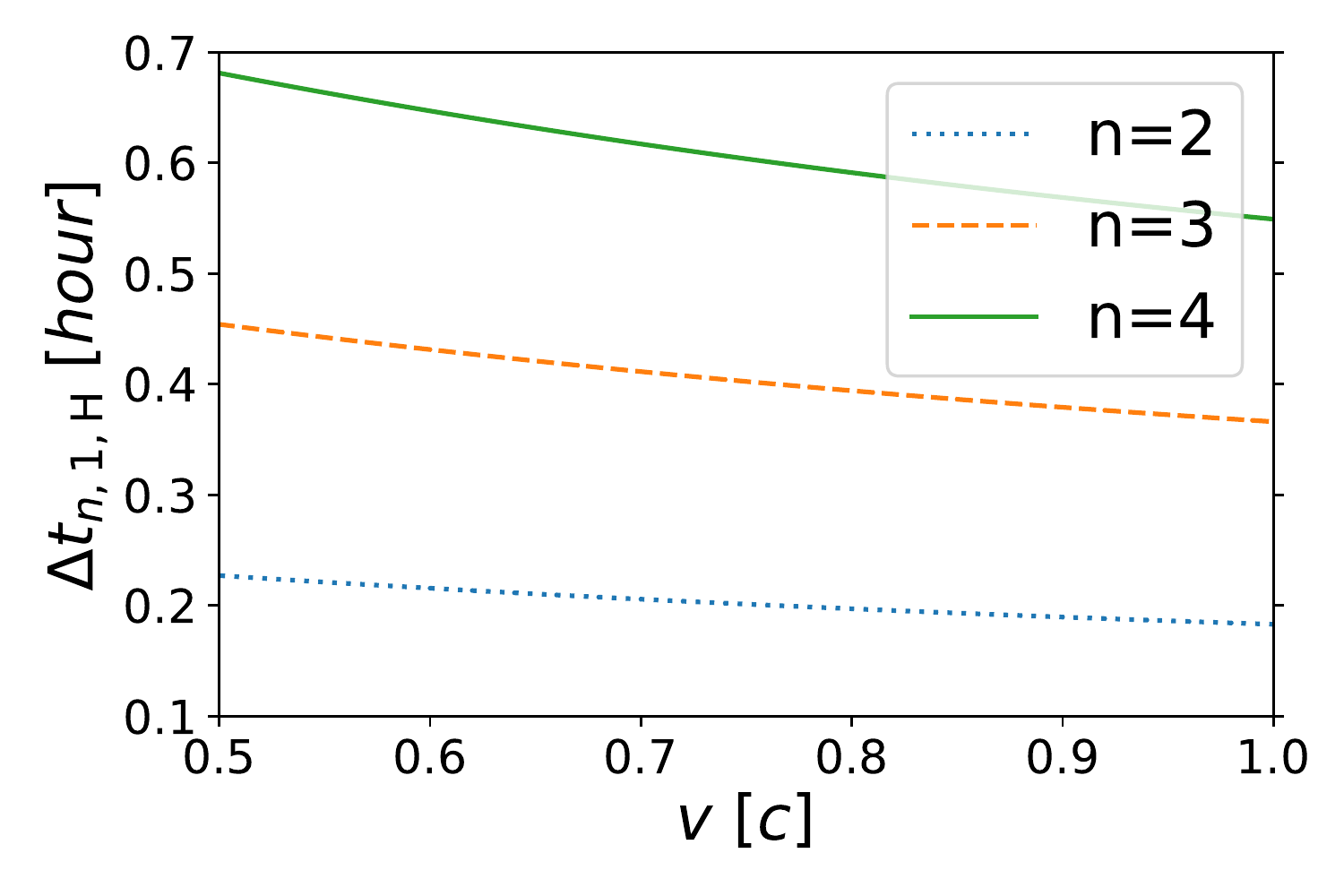}}\\
\subfigure[]{\label{fig:tdophi2} \includegraphics[width=0.48\textwidth]{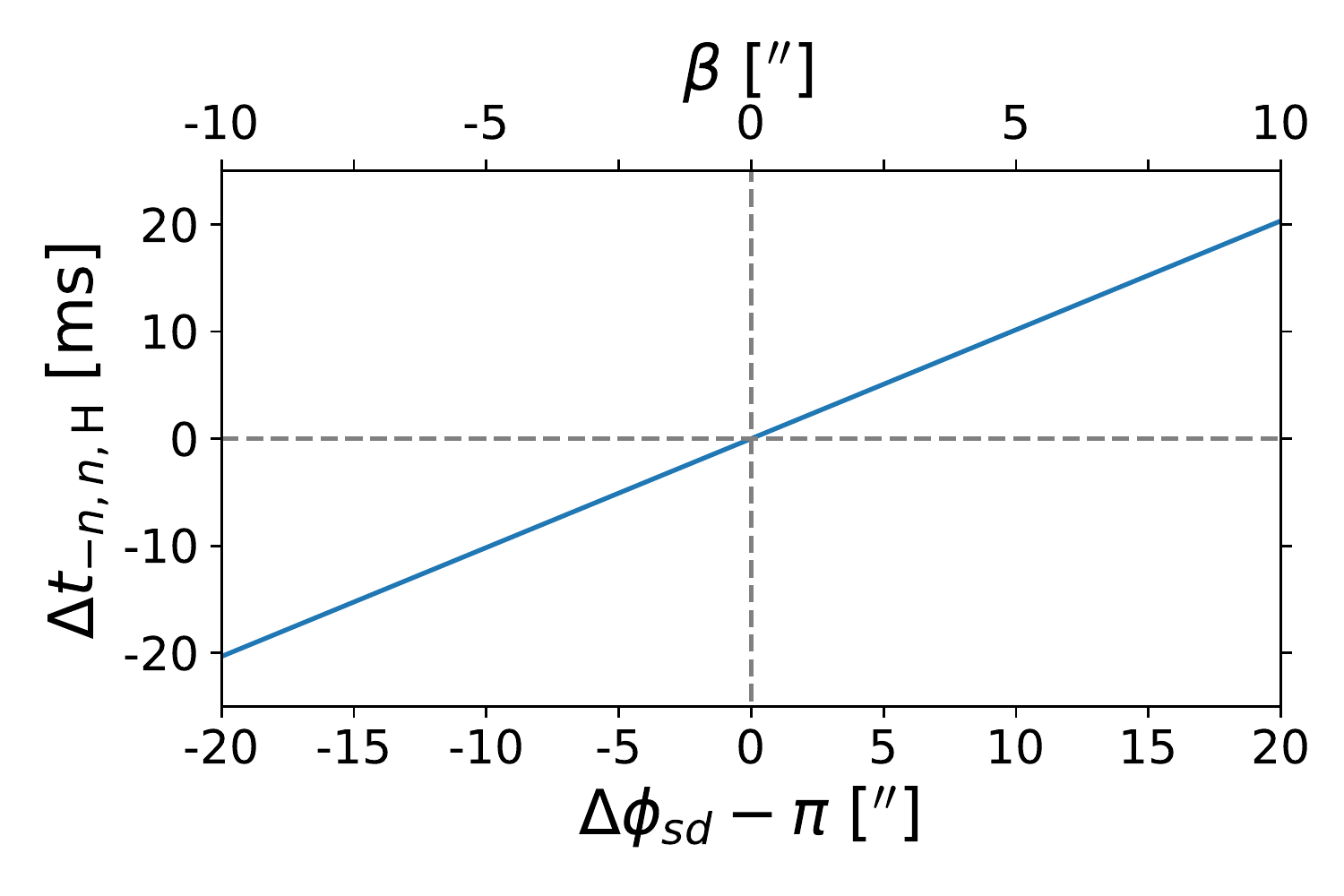}}
\caption{(a) The radius of the particle sphere as a function of spacetimes parameter $l$ and signal velocity $v$. Time delay between signals from same side but with different winding numbers as a function of: (a) $l$ while holding $v=1$ and; (b) $v$ while holding $l/M=0.1$. In (d), the time delay as a function of $\Delta \phi_{sd}$ for signals from opposite sides but with the same absolute wind number, with parameter $l/M=0.1,~v=1$ and $r_s=r_d=r_{\mathrm{Sgr~A*}}$ and $n>0$.  } \label{fig:tdolvphi}
\end{figure}
\end{center}

Although the truncated $t_m(r_s, r_d, b)$ in Eq. \eqref{eq:hayttm} approximate the true value pretty well in Fig. \ref{fig:ttobrsd} in the SFL, the location of the source and especially the detector used there is much smaller than their practical values. When $r_d/M$ (and $r_s/M$) is as large as in any practical GL, a numerical study shows that to achieve the same accuracy in the total travel time as in Fig. \ref{fig:ttobrsd}, the truncation order of $t_m(r_s, r_d, b)$ would be formidably high, even in the SFL.  Fortunately, this will not affect the accuracy of the time delay \eqref{eq:tdtotapp} because in the SFL, the time delay  between different trajectories mainly happens when the signal is very close to the particle sphere, for which part the total time \eqref{eq:hayttm} is a very good approximation. In other words, for any two trajectories, the travel times corresponding to the parts from large $r_d$ or $r_s$ to some small radius -- below which the time delay happens -- cancel out. This indeed leaves the time delay formula \eqref{eq:tdtotapp} very accurate even we truncated at a relatively low order, as can be seen from Fig. \ref{fig:tdolvphi}. 

The effects of the spacetime charge $l$ and signal velocity $v$, as well as $\Delta\phi_{sd}$ when $\sign(n)\neq\sign(m)$ are shown in Fig. \ref{fig:tdolvphi}(b)-(d) by assuming that the Sgr A* supermassive BH is a Hayward BH. To help understanding these effects, we first plot the particle sphere radius $r_c$ as a function of $l/M$ and $v$ in Fig. \ref{fig:tdolvphi}(a). It is seen that for any particular signal velocity, as the charge $l$ increases from 0 to $l_c=4/(3\sqrt{3}M)$, $r_c$ decreases monotonically from its Schwarzschild value to a minimal value. In particular,  for light signal, this is from $3M$ to about $2.65M$.  For fixed $l$ and decreasing  $v$ on the other hand, $r_c$ increases from its light signal value to a larger but still finite radius. Comparing to  the Reissner-Nordstrom spacetime, we see that the effects of $l$ and $v$ on $r_c$ are qualitatively similar to those of the electrostatic charge and particle velocity in that spacetime  \cite{Pang:2018jpm}. 

Then for the time delay $\Delta t_{n,m,\mathrm{H}}$ in Eq. \eqref{eq:dthayward}, it is clear that as $l$ increases to $l_c$,  $r_c$ decreases and changes of terms in both the denominator and numerator of the first factor cancel to a large extent. Therefore this factor has a very minimal variation. Similar trend happens for the second term and therefore the time delay depends very weakly on $l$, as can be seen from Fig. \ref{fig:tdolvphi}(b). In the entire range of $l$ from 0 to $l_c$, the time delay  $\Delta t_{n,m,\mathrm{H}}$ for light signals from the same side of the lens changes from $0.183|n-m|$ [hour]  to $0.163|n-m|$ [hour], a difference of $7.20\times 10^1$ [second] per loop.  For the signal velocity,  it is seen that as $v/c$ decreases from 1 to 0.5, the time delay increases from $0.183|n-m|$ [hour]  to $0.227|n-m|$ [hour], a difference of $2.64\times 10^2$ [second] per loop. From the time measurement perspective, fortunately both these two changes per loop are well resolvable, as long as the characteristic time scale of the source event is not larger than these values. Events satisfying this certainly include typical supernova explosion and binary BH/Neutron star merger, whose characteristic time scale are maximally $\sim 10$ seconds and $0.1\sim 1$ second respectively. Therefore their observation might be used to constrain the value of parameter $l$ to good accuracy. 

While to constrain signal velocity, for typical supernova neutrinos with energy at the $\mathcal{O}(10)$ [MeV] level, their velocity can only deviate from light speed by $1.25\times 10^{-17}c$ at most \cite{Zyla:2020zbs} and speed of GW has already been constrained to be within $3\times 10^{-15}c$ from light speed \cite{TheLIGOScientific:2017qsa}. For these velocities, using plot Fig. \ref{fig:tdolvphi}(c) one can estimate the maximal difference between the time delays of these timelike signals and that of the light signal. For the former, this difference is only about $6.60\times 10^{-15}$ [s] and for the latter $1.58\times 10^{-12}$ [s]. These are not only much shorter than characteristic time scale of supernova and binary merger, but also smaller by several orders than the corresponding detector resolution ($\mathcal{O}(1)$ [ns] for neutrino observatories and $\mathcal{O}(1)$ [ms] for GW detectors \cite{TheLIGOScientific:2017qsa}). Therefore using time delay caused by Sgr A* in the SFL to constrain speed of such ultra-relativistic signals seems not likely.

Finally, for the case of signals from opposite sides of the lens, i.e., $\sign(n)\neq\sign(m)$, the dependence of the time delay on $\Delta\phi_{sd}$ is shown in Fig. \ref{fig:tdolvphi}(d). When $n=-m$ and $\Delta\phi_{sd}=\pi$, i.e., the source is perfectly aligned along the observer-lens axis and the trajectories from two sides have an mirror symmetry, then clearly we should have $\Delta t_{-m,m,\mathrm{H}}(\Delta\phi_{sd}=\pi)=0$, as shown in Fig. \ref{fig:tdophi2}.  Unlike GL in the weak field limit, angle $\Delta \phi_{sd}$ dose not need to be very close to $\pi$ if GL in the SFL can really be observed in the future. As $\Delta\phi_{sd}$ deviates from $\pi$, the time delay $\Delta t_{-1,1,\mathrm{H}}$ becomes linear to $(\Delta\phi_{sd}-\pi)$. The arrival time of the each series of images from one side of the lens will form an arithmetic sequence which is equivalent to $\Delta t_{n,1,\mathrm{H}}~(n=2,3,\cdots)$ or $\Delta t_{n,-1,\mathrm{H}}~(n=-2,-3,\cdots)$. The two sequences from two sides will have a relative shift $\Delta t_{-n,n,\mathrm{H}}$ that is linear to $(\Delta\phi_{sd}-\pi)$ too, as shown in Fig. \ref{fig:tdolvphi}\ref{fig:tdophi2}. Because of this relation, it is seen that for a given characteristic time scale of the source event or observatory time resolution (whichever is larger), there exist a minimal $\Delta\phi_{sd}-\pi$ that allows the two sequences to be temporally separated. Taking 2 [ms] for the time resolution of GW signal as an example, then it is seen that for the two sequence to be separated by this interval, $|\Delta\phi_{sd}-\pi| $ has to be larger than 2 [$''$]. 
On the other hand, if GL of an event is to be both observed in the weak field limit and temporally resolved in the SFL, then since GL in the weak field limit are usually observed for $\beta\lesssim 10$ [$''$] \cite{gldatabase} or roughly $\Delta\phi_{sd}\lesssim 20$ [$''$], Fig. \ref{fig:tdolvphi}(d) implies that both the characteristic time scale of the event and the observatory time resolution have to be smaller than 20 [ms].  Therefore to temporally resolve the two sequences from two sides certainly imposes stringent requirement on the time measurement uncertainty of observatories, e.g., the GRB measurement uncertainty which is current about 50 [ms]  \cite{Monitor:2017mdv}.

\section{Conclusion and discussion}

In this work we proposed a perturbative method to compute the total travel time $t$ and time delay $\Delta t$ in the SFL in SSS spacetimes for signal with arbitrary asymptotic velocity. The total travel time takes a simple quasi-series form
\be
t=\sum_{k=0}^\infty \lsb C_k\ln a +D_k\rsb a^k,
\ee
where $a=\lb 1-\frac{b_c}{b}\rb $ and coefficients $C_k$ and $D_k$ can be expressed as rational functions of the metric expansion coefficients at the particle sphere radius.
In the SFL, the leading contribution to $\Delta t$ is given by the $\ln a$ term. Using impact parameter corresponding to each relativistic image, we were able to show that $\Delta t$ is given by Eq. \eqref{eq:tdtotapp}. This result allows an intuitive and yet quantitatively precise understanding: to the leading order of $a$, the time delay is given by the circumference of the particle sphere divided by the local velocity of the signal and then multiplied by the winding number difference and the redshift factor from the particle sphere to the far away detector. 
 
We applied the results of $t$ and $\Delta t$ to the Hayward BH spacetime. The correctness of the total travel time is verified by truncating the series to different orders. The time delay in this case is found in Eq. \eqref{eq:haytt}. To understand it, we first studied the dependence of the particle sphere radius $r_c$ on the spacetime charge parameter $l$ and signal velocity $v$. It is found that as $l$ increases or $v$ decreases, $r_c$ decreases or increases respectively. Assuming the Sgr A* central BH is a Hayward BH, we were able to compute $\Delta t$ between images on the same side and opposite sides. It is found that as $l$ increases from 0 to its critical value, the time delay per loop can vary by $7.2\times 10^1$ [s], which is well within the time resolution of typical light, neutrino or GW observatories. Therefore measuring $\Delta t$ caused by the Sgr A* BH will constrain its charge $l$ very well. On the other hand, for the supernova neutrinos or GW whose velocities deviate from that of the light by $10^{-15}$ or less, measuring difference between $\Delta t$'s of different signals would not further constrain their velocities, because of the time measurement accuracy/characteristic time of the corresponding events are larger by a few orders. 

Regarding directions to extend the current work, the first and most straightforward one is to extend the perturbative method to the case of the equatorial plane in the stationary and axialsymmetric spacetime. Based on the weak field limit experience \cite{Liu:2020mkf}, we expect that the spin parameter would play a non-trivial role in affecting the time delay between relativistic images with different winding directions.  A more interesting extension is to apply the method to the time delay of asymptotically non-flat spacetimes. From the quasi-series \eqref{eq:totaltaf} for the total time, we saw that the coefficients $C_k$ and $D_k$ are completely determined by the behavior of the metric functions around the particle sphere. Although GL in the weak field limit in these spacetimes is often problematic to study due to the difficulty to take the infinite radius limit, the metric functions behave normally at small radius and therefore the SFL can still be taken. We are pursuing along these directions.

\acknowledgments
We thank Dr. Nan Yang and Mr. Ke Huang for valuable discussions. This work is supported by the NNSF China 11504276.

\appendix

\section{High order $y_{n,0}$ and integration formulas} \label{app:hoyz}

The integrals in Eq. \eqref{eq:ttgenexpform} can be worked out as the following: for even $n=2k$,
\begin{subequations} \label{eq:ttgenintm}
\begin{align}
&  \int_{a}^{a_i}\frac{\xi^k}{\sqrt{\xi-a}} \dd \xi=\sum_{j=0}^{k} \frac{2 C_{j}^{k}a^{k-j}\lb a_i-a\rb^{j+1/2}}{2j+1}\text{ and for odd }n=2k-1,  \label{eq:ttgenintmek} \\
&\int_{a}^{a_i}\frac{\xi^{k-1/2}}{\sqrt{\xi-a}} \dd \xi= \frac{a^kC_{2k}^k}{4^k}\lsb 2\ln\lb \sqrt{\frac{a_i}{a}}+\sqrt{\frac{a_i}{a}-1}\rb + \sum_{j=1}^{k}\frac{4^j}{jC_{2j}^{j}}\lb \frac{a_i}{a}\rb^j\sqrt{1-\frac{a}{a_i}}\rsb, \label{eq:ttgenintmok}
\end{align}
\end{subequations}
where $k$ are non-negative integers.

The high order coefficient of $y_{n,0}$ in Eq. \eqref{eqs:gencoeyaz} can be worked out with the help of a symbolic computation tool. Here we will only give one more order, i.e.,
\begin{align}
y_{1,0} =& \frac{\sqrt{ b_0  c_0} }{4 \sqrt{2}v a_0^3 b_0^2 c_0^2 T_2^{7/2}} \lcb a_0^2 \lsb b_0^2 \left( -2 b_c^2 \left(6 c_0 c_1 T_2 T_3+ c_0 \left(3 c_0 \left(4  T_2 T_4-5 T_3^2\right) \right. \right. \right. \right. \right. \nonumber \\
  & \left. \left. \left. -4 c_2  T_2^2\right)+ c_1^2 T_2^2\right)+ 12 c_0^2  T_2^3\right) + 4 b_0 c_0 b_c^2 T_2 (-3 b_1 c_0 T_3+ b_1 c_1 T_2+2 b_2 c_0  T_2) \nonumber \\
  & \left. - 2 b_1^2 c_0^2 b_c^2 T_2^2\rsb - 8 a_0 b_0 c_0 b_c^2 T_2 \lsb a_1 (-3 b_0 c_0 T_3+ b_0 c_1 T_2+ b_1 c_0 T_2)+ 2 a_2 b_0 c_0 T_2 \rsb \nonumber \\
  & \left. +16 a_1^2  b_0^2 c_0^2 b_c^2 T_2^2 \rcb.
\end{align}
where
\begin{align}
T_4=& \frac{1}{v^2 a_0} \lsb c_4 \lb 1 - a_0 + a_0 v^2 \rb - \frac{ \lb a_1 c_3 + a_2 c_2 + a_3 c_1 + a_4 c_0 \rb}{a_0} \right. \nonumber \\
   & \left. + \frac{\lb a_1^2 c_2 + 2 a_1 a_2 c_1 + 2 a_1 a_3 c_0 + a_2^2 c_0 \rb}{a_0^2} - \frac{\lb a_1^3 c_1 + 3 a_1^2 a_2 c_0 \rb}{a_0^3} + \frac{a_1^4 c_0}{a_0^4} \rsb.
\end{align}


\begin{thebibliography}{99}

\bibitem{Dyson:1920cwa} F.~W.~Dyson, A.~S.~Eddington and C.~Davidson, 
Phil. Trans. Roy. Soc. Lond. A \textbf{220} (1920), 291-333 
doi:10.1098/rsta.1920.0009 

\bibitem{Bradac:2006er} M.~Bradac, D.~Clowe, A.~H.~Gonzalez, P.~Marshall, W.~Forman, C.~Jones, M.~Markevitch, S.~Randall, T.~Schrabback and D.~Zaritsky,
Astrophys. J. \textbf{652} (2006), 937-947
doi:10.1086/508601
[arXiv:astro-ph/0608408 [astro-ph]].

\bibitem{Treu:2010uj} T.~Treu,
Ann. Rev. Astron. Astrophys. \textbf{48} (2010), 87-125
doi:10.1146/annurev-astro-081309-130924
[arXiv:1003.5567 [astro-ph.CO]].

\bibitem{Suyu:2012aa} S.~H.~Suyu, M.~W.~Auger, S.~Hilbert, P.~J.~Marshall, M.~Tewes, T.~Treu, C.~D.~Fassnacht, L.~V.~E.~Koopmans, D.~Sluse and R.~D.~Blandford, \textit{et al.}
Astrophys. J. \textbf{766} (2013), 70
doi:10.1088/0004-637X/766/2/70
[arXiv:1208.6010 [astro-ph.CO]].

\bibitem{Guo:2009fn} Q.~Guo, S.~White, C.~Li and M.~Boylan-Kolchin,
Mon. Not. Roy. Astron. Soc. \textbf{404} (2010), 1111
doi:10.1111/j.1365-2966.2010.16341.x
[arXiv:0909.4305 [astro-ph.CO]].

\bibitem{Zu:2015cpa} Y.~Zu and R.~Mandelbaum,
Mon. Not. Roy. Astron. Soc. \textbf{454} (2015) no.2, 1161-1191
doi:10.1093/mnras/stv2062
[arXiv:1505.02781 [astro-ph.CO]].

\bibitem{Wechsler:2018pic} R.~H.~Wechsler and J.~L.~Tinker,
Ann. Rev. Astron. Astrophys. \textbf{56} (2018), 435-487
doi:10.1146/annurev-astro-081817-051756
[arXiv:1804.03097 [astro-ph.GA]].

\bibitem{Morgan:2010xf} C.~W.~Morgan, C.~S.~Kochanek, N.~D.~Morgan and E.~E.~Falco,
Astrophys. J. \textbf{712} (2010), 1129-1136
doi:10.1088/0004-637X/712/2/1129
[arXiv:1002.4160 [astro-ph.CO]].

\bibitem{Diemer:2014xya} B.~Diemer and A.~V.~Kravtsov,
Astrophys. J. \textbf{789} (2014), 1
doi:10.1088/0004-637X/789/1/1
[arXiv:1401.1216 [astro-ph.CO]].

\bibitem{Kelly:2014mwa} P.~L.~Kelly {\it et al.},
  Science {\bf 347}, 1123 (2015)
  doi:10.1126/science.aaa3350
  [arXiv:1411.6009 [astro-ph.CO]].

\bibitem{Goobar:2016uuf} A.~Goobar {\it et al.},
  Science {\bf 356}, 291 (2017)
  doi:10.1126/science.aal2729
  [arXiv:1611.00014 [astro-ph.CO]].

\bibitem{Fan:2016swi} X.~L.~Fan, K.~Liao, M.~Biesiada, A.~Piorkowska-Kurpas and Z.~H.~Zhu,
  Phys.\ Rev.\ Lett.\  {\bf 118}, no. 9, 091102 (2017)

\bibitem{Hirata:1987hu} K.~Hirata {\it et al.} [Kamiokande-II Collaboration],
Phys.\ Rev.\ Lett.\ {\bf 58}, 1490 (1987).

\bibitem{Bionta:1987qt} R.~M.~Bionta {\it et al.},
Phys.\ Rev.\ Lett.\ {\bf 58}, 1494 (1987).

\bibitem{IceCube:2018dnn} M.~G.~Aartsen {\it et al.} [IceCube and Fermi-LAT and MAGIC and AGILE and ASAS-SN and HAWC and H.E.S.S. and INTEGRAL and Kanata and Kiso and Kapteyn and Liverpool Telescope and Subaru and Swift NuSTAR and VERITAS and VLA/17B-403 Collaborations],
  Science {\bf 361}, no. 6398, eaat1378 (2018)

\bibitem{IceCube:2018cha} M.~G.~Aartsen {\it et al.} [IceCube Collaboration],
  Science {\bf 361}, no. 6398, 147 (2018)

\bibitem{Abbott:2016blz} B.~P.~Abbott {\it et al.} [LIGO Scientific and Virgo Collaborations],
  Phys.\ Rev.\ Lett.\  {\bf 116}, no. 6, 061102 (2016)
  doi:10.1103/PhysRevLett.116.061102
  [arXiv:1602.03837 [gr-qc]].

\bibitem{Abbott:2016nmj} B.~P.~Abbott {\it et al.} [LIGO Scientific and Virgo Collaborations],
  Phys.\ Rev.\ Lett.\  {\bf 116}, no. 24, 241103 (2016)
  doi:10.1103/PhysRevLett.116.241103
  [arXiv:1606.04855 [gr-qc]].

\bibitem{Abbott:2017oio} B.~P.~Abbott {\it et al.} [LIGO Scientific and Virgo Collaborations],
  Phys.\ Rev.\ Lett.\  {\bf 119}, no. 14, 141101 (2017)
  doi:10.1103/PhysRevLett.119.141101
  [arXiv:1709.09660 [gr-qc]].

\bibitem{TheLIGOScientific:2017qsa} B.~P.~Abbott {\it et al.} [LIGO Scientific and Virgo Collaborations],
  Phys.\ Rev.\ Lett.\  {\bf 119}, no. 16, 161101 (2017)
  doi:10.1103/PhysRevLett.119.161101
  [arXiv:1710.05832 [gr-qc]].

\bibitem{Eberhardt:2017vce} B.~Eberhardt,

\bibitem{Djurcic:2015vqa} Z.~Djurcic \textit{et al.} [JUNO],
[arXiv:1508.07166 [physics.ins-det]].

\bibitem{Bozza:2003cp} V.~Bozza and L.~Mancini,
Gen. Rel. Grav. \textbf{36} (2004), 435-450
doi:10.1023/B:GERG.0000010486.58026.4f
[arXiv:gr-qc/0305007 [gr-qc]].

\bibitem{Eiroa:2004gh} E.~F.~Eiroa,
Phys. Rev. D \textbf{71} (2005), 083010
doi:10.1103/PhysRevD.71.083010
[arXiv:gr-qc/0410128 [gr-qc]].

\bibitem{Eiroa:2005ag} E.~F.~Eiroa,
  Phys.\ Rev.\ D {\bf 73}, 043002 (2006)
  doi:10.1103/PhysRevD.73.043002
  [gr-qc/0511065].

\bibitem{Eiroa:2013nra} E.~F.~Eiroa and C.~M.~Sendra,
  Phys.\ Rev.\ D {\bf 88}, no. 10, 103007 (2013)
  doi:10.1103/PhysRevD.88.103007
  [arXiv:1308.5959 [gr-qc]].

\bibitem{Cavalcanti:2016mbe} R.~T.~Cavalcanti, A.~G.~da Silva and R.~da Rocha,
Class. Quant. Grav. \textbf{33} (2016) no.21, 215007
doi:10.1088/0264-9381/33/21/215007
[arXiv:1605.01271 [gr-qc]].

\bibitem{Zhao:2017cwk} S.~S.~Zhao and Y.~Xie,
Eur. Phys. J. C \textbf{77} (2017) no.5, 272
doi:10.1140/epjc/s10052-017-4850-5
[arXiv:1704.02434 [gr-qc]].

\bibitem{Wang:2019cuf} C.~Y.~Wang, Y.~F.~Shen and Y.~Xie,
JCAP \textbf{04} (2019), 022
doi:10.1088/1475-7516/2019/04/022
[arXiv:1902.03789 [gr-qc]].

\bibitem{Sahu:2015dea} S.~Sahu, K.~Lochan and D.~Narasimha,
Phys. Rev. D \textbf{91} (2015), 063001
doi:10.1103/PhysRevD.91.063001
[arXiv:1502.05619 [gr-qc]].

\bibitem{Man:2015iga} J.~Man and H.~Cheng,
Phys. Rev. D \textbf{92} (2015) no.2, 024004
doi:10.1103/PhysRevD.92.024004
[arXiv:1205.4857 [gr-qc]].

\bibitem{Majumdar:2005ba} A.~S.~Majumdar and N.~Mukherjee,
Int. J. Mod. Phys. D \textbf{14} (2005), 1095
doi:10.1142/S0218271805006948
[arXiv:astro-ph/0503473 [astro-ph]].

\bibitem{Gyulchev:2008ff} G.~N.~Gyulchev and S.~S.~Yazadjiev,
Phys. Rev. D \textbf{78} (2008), 083004
doi:10.1103/PhysRevD.78.083004
[arXiv:0806.3289 [gr-qc]].

\bibitem{Sahu:2013uya} S.~Sahu, M.~Patil, D.~Narasimha and P.~S.~Joshi,
Phys. Rev. D \textbf{88} (2013), 103002
doi:10.1103/PhysRevD.88.103002
[arXiv:1310.5350 [gr-qc]].

\bibitem{Jia:2020huang} J.~Jia and K.~Huang,
[arXiv:2011.08084 [gr-qc]].

\bibitem{Liu:2020wcu} H.~Liu and J.~Jia,
[arXiv:2006.03542 [gr-qc]].

\bibitem{Hayward:2005gi} S.~A.~Hayward, 

\bibitem{Pang:2018jpm} X.~Pang and J.~Jia,
Class. Quant. Grav. \textbf{36} (2019) no.6, 065012
doi:10.1088/1361-6382/ab0512
[arXiv:1806.04719 [gr-qc]].

\bibitem{Zyla:2020zbs} P.~A.~Zyla \textit{et al.} [Particle Data Group],
PTEP \textbf{2020}, no.8, 083C01 (2020)
doi:10.1093/ptep/ptaa104

\bibitem{Monitor:2017mdv} B.~P.~Abbott {\it et al.} [LIGO Scientific and Virgo and Fermi-GBM and INTEGRAL Collaborations],
  Astrophys.\ J.\  {\bf 848}, no. 2, L13 (2017)
  doi:10.3847/2041-8213/aa920c
  [arXiv:1710.05834 [astro-ph.HE]].

\bibitem{Bozza:2009yw} V.~Bozza,
Gen. Rel. Grav. \textbf{42} (2010), 2269-2300
doi:10.1007/s10714-010-0988-2
[arXiv:0911.2187 [gr-qc]].

\bibitem{Cunha:2018acu} P.~V.~P.~Cunha and C.~A.~R.~Herdeiro,
Gen. Rel. Grav. \textbf{50} (2018) no.4, 42
doi:10.1007/s10714-018-2361-9
[arXiv:1801.00860 [gr-qc]].

\bibitem{Gralla:2019xty} S.~E.~Gralla, D.~E.~Holz and R.~M.~Wald,
Phys. Rev. D \textbf{100} (2019) no.2, 024018
doi:10.1103/PhysRevD.100.024018
[arXiv:1906.00873 [astro-ph.HE]].

\bibitem{Akiyama:2019cqa} K.~Akiyama {\it et al.} [Event Horizon Telescope Collaboration],
  Astrophys.\ J.\  {\bf 875}, no. 1, L1 (2019)
  doi:10.3847/2041-8213/ab0ec7
  [arXiv:1906.11238 [astro-ph.GA]].

\bibitem{Akiyama:2019brx} K.~Akiyama \textit{et al.} [Event Horizon Telescope],
Astrophys. J. Lett. \textbf{875} (2019) no.1, L2
doi:10.3847/2041-8213/ab0c96
[arXiv:1906.11239 [astro-ph.IM]].

\bibitem{Akiyama:2019sww} K.~Akiyama \textit{et al.} [Event Horizon Telescope],
Astrophys. J. Lett. \textbf{875} (2019) no.1, L3
doi:10.3847/2041-8213/ab0c57
[arXiv:1906.11240 [astro-ph.GA]].

\bibitem{Akiyama:2019bqs} K.~Akiyama \textit{et al.} [Event Horizon Telescope],
Astrophys. J. Lett. \textbf{875} (2019) no.1, L4
doi:10.3847/2041-8213/ab0e85
[arXiv:1906.11241 [astro-ph.GA]].

\bibitem{Akiyama:2019fyp} K.~Akiyama \textit{et al.} [Event Horizon Telescope],
Astrophys. J. Lett. \textbf{875} (2019) no.1, L5
doi:10.3847/2041-8213/ab0f43
[arXiv:1906.11242 [astro-ph.GA]].

\bibitem{Akiyama:2019eap} K.~Akiyama \textit{et al.} [Event Horizon Telescope],
Astrophys. J. Lett. \textbf{875} (2019) no.1, L6
doi:10.3847/2041-8213/ab1141
[arXiv:1906.11243 [astro-ph.GA]].

\bibitem{Bozza:2002zj} V.~Bozza,
Phys. Rev. D \textbf{66} (2002), 103001
doi:10.1103/PhysRevD.66.103001
[arXiv:gr-qc/0208075 [gr-qc]].

\bibitem{Virbhadra:2008ws} K.~S.~Virbhadra,
Phys. Rev. D \textbf{79} (2009), 083004
doi:10.1103/PhysRevD.79.083004
[arXiv:0810.2109 [gr-qc]].

\bibitem{GBM:2017lvd} B.~P.~Abbott \textit{et al.} [LIGO Scientific, Virgo, Fermi GBM, INTEGRAL, IceCube, AstroSat Cadmium Zinc Telluride Imager Team, IPN, Insight-Hxmt, ANTARES, Swift, AGILE Team, 1M2H Team, Dark Energy Camera GW-EM, DES, DLT40, GRAWITA, Fermi-LAT, ATCA, ASKAP, Las Cumbres Observatory Group, OzGrav, DWF (Deeper Wider Faster Program), AST3, CAASTRO, VINROUGE, MASTER, J-GEM, GROWTH, JAGWAR, CaltechNRAO, TTU-NRAO, NuSTAR, Pan-STARRS, MAXI Team, TZAC Consortium, KU, Nordic Optical Telescope, ePESSTO, GROND, Texas Tech University, SALT Group, TOROS, BOOTES, MWA, CALET, IKI-GW Follow-up, H.E.S.S., LOFAR, LWA, HAWC, Pierre Auger, ALMA, Euro VLBI Team, Pi of Sky, Chandra Team at McGill University, DFN, ATLAS Telescopes, High Time Resolution Universe Survey, RIMAS, RATIR and SKA South Africa/MeerKAT],
Astrophys. J. Lett. \textbf{848} (2017) no.2, L12
doi:10.3847/2041-8213/aa91c9
[arXiv:1710.05833 [astro-ph.HE]].

\bibitem{Eiroa:2008ks} E.~F.~Eiroa and G.~E.~Romero,
Phys. Lett. B \textbf{663} (2008), 377-381
doi:10.1016/j.physletb.2008.04.016
[arXiv:0802.4251 [astro-ph]].

\bibitem{Muller:2008zza} T.~Muller,
Phys. Rev. D \textbf{77} (2008), 044043
doi:10.1103/PhysRevD.77.044043

\bibitem{Crisnejo:2018uyn} G.~Crisnejo and E.~Gallo,
Phys. Rev. D \textbf{97} (2018) no.12, 124016
doi:10.1103/PhysRevD.97.124016
[arXiv:1804.05473 [gr-qc]].

\bibitem{Takahashi:2016jom} R.~Takahashi,
Astrophys. J. \textbf{835} (2017) no.1, 103
doi:10.3847/1538-4357/835/1/103
[arXiv:1606.00458 [astro-ph.CO]].

\bibitem{Baker:2016reh} T.~Baker and M.~Trodden,
Phys. Rev. D \textbf{95} (2017) no.6, 063512
doi:10.1103/PhysRevD.95.063512
[arXiv:1612.02004 [astro-ph.CO]].

\bibitem{Jia:2015zon} X.~Liu, J.~Jia and N.~Yang,
Class. Quant. Grav. \textbf{33} (2016) no.17, 175014
doi:10.1088/0264-9381/33/17/175014
[arXiv:1512.04037 [gr-qc]].

\bibitem{Jia:2019hih} J.~Jia and H.~Liu,
  Phys.\ Rev.\ D {\bf 100}, no. 12, 124050 (2019)
  doi:10.1103/PhysRevD.100.124050
  [arXiv:1906.11833 [gr-qc]].

\bibitem{Li:2019qyb} Z.~Li and J.~Jia,
Eur. Phys. J. C \textbf{80} (2020) no.2, 157
doi:10.1140/epjc/s10052-020-7665-8
[arXiv:1912.05194 [gr-qc]].

\bibitem{Jia:2020xbc} J.~Jia,
Eur. Phys. J. C \textbf{80} (2020) no.3, 242
doi:10.1140/epjc/s10052-020-7796-y
[arXiv:2001.02038 [gr-qc]].

\bibitem{Duan:2020tsq} Y.~Duan, W.~Hu, K.~Huang and J.~Jia,
Class. Quant. Grav. \textbf{37} (2020) no.14, 145004
doi:10.1088/1361-6382/ab852c
[arXiv:2001.03777 [gr-qc]].

\bibitem{Huang:2020trl} K.~Huang and J.~Jia,
JCAP \textbf{08} (2020), 016
doi:10.1088/1475-7516/2020/08/016
[arXiv:2003.08250 [gr-qc]].

\bibitem{Liu:2020mkf} H.~Liu and J.~Jia,
Eur. Phys. J. C \textbf{80} (2020) no.10, 932
doi:10.1140/epjc/s10052-020-08496-5
[arXiv:2006.11125 [gr-qc]].

\bibitem{gldatabase}
C.S. Kochanek, E.E. Falco, C. Impey, J. Lehar, B. McLeod, H.-W. Rix, https://www.cfa.harvard.edu/castles/
\end{thebibliography}
\end{document}